\documentstyle{l-aa}
\begin{document}

\thesaurus{02(12.03.1; 12.03.3; 03.13.6)}

\title{Peak statistics on COBE maps }
\author{R.~Fabbri\inst{1} \and S.~Torres \inst{2}}

\offprints{R.~Fabbri}

\institute{Dipartimento di Fisica dell'Universit\`{a},
 Sezione di Fisica Superiore,
 Via S. Marta 3,  I-50139 Firenze,  Italy
 \and Observatorio Astron\'omico Nacional, Universidad Nacional de Colombia,
Bogot\`a, Colombia}

\date{Received ; accepted }

\maketitle

\begin{abstract}
We perform the stastistics of temperature maxima and minima in COBE-DMR
2-year maps. For power-law spectra the surface distribution of peaks implies
an amplitude consistent with more conventional analyses of COBE data (for
instance, we get $Q_{{\rm rms-PS}}=17\pm 3\ \mu $K for a spectral index $%
n=1),$ but not with the measured quadrupole $Q_{{\rm rms}}=6\pm 3\ \mu $K.
This provides further support for the existence an infrared cutoff in the
cosmic spectrum.
\end{abstract}

\thesaurus{02(12.03.1; 12.03.3; 03.13.6)}

\offprints{R.Fabbri}
\institute{Dipartimento di Fisica dell'Universit\`{a},
 Sezione di Fisica Superiore,
 Via S. Marta 3,  I-50139 Firenze,  Italy
 \and Observatorio Astron\'omico Nacional, Universidad Nacional de Colombia,
Bogot\`a, Colombia}

\section{Introduction}

The COBE experiment (Bennet et al. 1992, 1994; Smoot et al. 1992) has
stimulated a considerable amount of work on cosmic structure. Current tests
usually exploit the angular correlation function and several harmonic
amplitudes of the sky temperature field (see e.g, Adams et al. 1992;
Kashlinsky 1992; Efstathiou et al. 1992; Kofmann et al. 1993; Gorski et al.
1994). However, several more tests have been suggested over the years for a
thorough investigation of the properties of the anisotropy field of the
cosmic background radiation. These often involve the distribution and
features of hot and cold spots, which can provide useful checks of the
Gaussian nature of the fluctuations (Sazhin 1985; Bond \& Efstathiou 1987;
Coles \& Barrow 1987; Coles 1988; Mart\'\i nez-Gonz\'alez \& Sanz 1989; Gott
et al. 1990). Measurable quantities include the number of spots $N_{{\rm iso}%
}$ defined by isotemperature contours, the spot boundary curvature or genus $%
G$, the spot excursion area, and so on. A first analysis of COBE-DMR maps
along these lines has been performed by Torres (1994). A definition of
spots independent of isotemperature contours considers local maxima and
minima of temperature (Bond \& Efstathiou 1987, Vittorio \& Juszkiewicz
1987), and is thereby not connected to the topological features of spots.
Differing from $N_{{\rm iso}},$ the dependence of the number of (positive
and/or negative) peaks $N_{{\rm peak}}$ on threshold is not universal for
Gaussian fields. This latter approach was adopted by Fabbri \& Natale (1993,
1995) in studies of the 2-dimensional distribution of extragalactic IRAS
sources, but has not yet been applied to the cosmic background radiation.

In this work we analyze the statistics of local maxima and minima in
COBE-DMR 2-year maps. We found that in this kind of analysis the detector
noise must be taken into account very carefully since $N_{{\rm peak}}$ is
sensitive also to high order harmonics where noise dominates (cf. Fabbri
1992). However, due to a highly nonlinear dependence of $N_{{\rm peak}}$ on
the harmonic strengths, the presence of structured signals in COBE maps
reduces its value below the level measured in pure noise maps. (The
identification of genuine peaks in the radiation temperature is not required
at all in our analysis.) We find that the distributions of positive and
negative peaks are mutually consistent, and the results from this statistics
agree with those of earlier tests. Therefore, we find no evidence for
non-Gaussian features in the fluctuations. More precisely, fitting Gaussian
power-law models of cosmic structure to the peak distribution we recover a
clear anticorrelation between the spectral index $n$ and the predicted rms
quadrupole $Q_{{\rm rms-PS}}$ (Seljak \& Bertschinger 1993; Smoot et al.
1994): We get $Q_{{\rm rms-PS}}=17\pm 3$ $\mu $K for $n=1$ and $14\pm 3$ $%
\mu $K for $n=1.5$, where the error bars include uncertainties deriving from
the treament of noise as well as cosmic variance. These numbers, altough
they agree with previous evaluations of the quadrupole from higher order
harmonics, are not consistent with its direct determination providing $Q_{%
{\rm rms}}=6\pm 3$ $\mu $K (Bennet et al. 1994). So the recently discovered
discrepancy is confirmed by the properties of the peak distribution, which
depend on the harmonic content of the angular distribution up to $\ell \sim
50.$

\section{The peak number test}

We analyzed the 2-year 53(A+B) DMR maps processed with a 2.9$^{\circ }$
smoothing (Wright et al. 1993) and dipole subtraction. Considering the
Northern and Southern hemispheres separately, we constructed two
pole-centered maps, each containing 12892 pixels, using the coordinate
transformation $\theta _1=2\sin \left[ \frac 12\left( \frac \pi 2-\left|
b\right| \right) \right] ,$ $\phi _1=l,$ with $b$ and $l$ the Galactic
coordinates. After masking low Galactic latitudes, $\left| b\right|
<20^{\circ },$ we were left with 8412 pixels per map. We then looked for
temperature peaks using the algorithms of Fabbri \& Natale (1993, 1995).

Table 1 gives the no-threshold numbers of peaks, both actually detected and
extrapolated to the entire sky and to the North and South hemispheres (2nd
and 3rd column, respectively). Figure 1 reports the extrapolated numbers vs.
a threshold factor $\nu $. This is the peak height normalized to the sky rms
fluctuation $C^{\frac 12}(0)$; for cold spots, $N_{{\rm peak}}$ gives the
number of minima below $-C^{\frac 12}(0)\nu $. For distributions of only
positive {\it or} negative peaks the statistical errors at 1-sigma
confidence levels are evaluated as $\left( N_{{\rm peak}}/f_{{\rm U}}\right)
^{\frac 12},$ with\/ $f_{{\rm U}}=0.652$ the unmasked fraction of the sky.
Within such error limits, we find no significant difference between the
distributions of positive and negative peaks; this result provides support
for the Gaussian nature of cosmic perturbations. In the figure we also
report the full-sky average number of positive and negative peaks. This
average will be compared with theoretical models below, because of the
smaller (by a factor $\sqrt{2})$ relative error.
\begin{table*}
\caption[ ]{No-threshold peak numbers \label{peaks}}
\begin{flushleft}
\begin{tabular}{llll}
\hline  \noalign {\smallskip}
     Peak set &  $(A+B)$ Maps  & $(A+B)$ Maps            & $(A-B)$ Maps\\
                 &  $|b| > 20^{\circ}$    &  Extrapolated$^{\dagger }$    & No
mask \\
\noalign {\smallskip}
\hline \noalign {\smallskip}
      Hot, North        &  50       &  $76.6 \pm 10.8 $ &  95   \\
      Cold, North       &  47       &  $72.0\pm 10.5  $ &  99  \\
      Hot, South        &  48       &  $73.6 \pm 10.6 $ &  105   \\
      Cold, South       &  53       &  $81.2 \pm 11.2 $ &  97  \\
      Hot, total          & 98      &  $ 150.2 \pm 15.2 $ & 200  \\
      Cold, total        & 100     &  $ 153.3 \pm 15.3 $ & 196  \\
      Hot/Cold Average           & 99     &  $ 151.7  \pm 10.8 $  & 198 \\
\noalign {\smallskip}
\hline
\end{tabular}

$^{\dagger }$To full sky or hemispheres, including masked regions
\end {flushleft}
\end{table*}
\begin{figure}
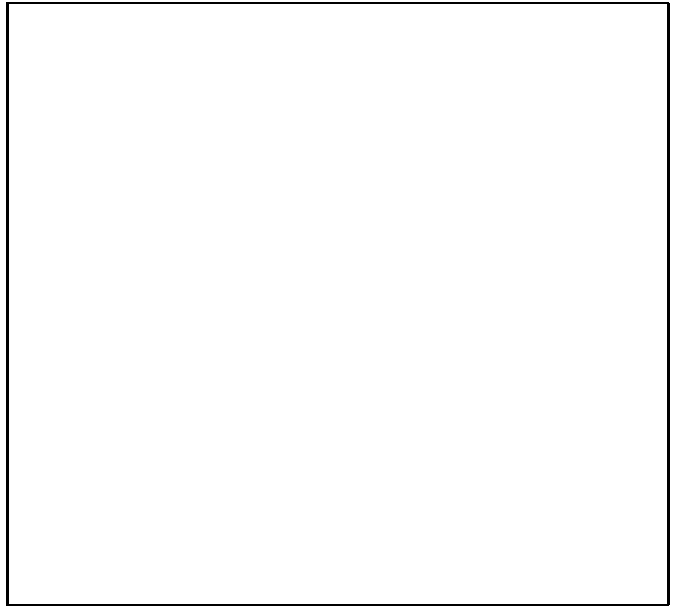

\picplace{8.0cm}
\caption[ ] {The peak number $ N_{\rm peak}$ vs. threshold in
COBE-DMR maps. Open
squares (triangles) denote positive (negative) peaks. The four lowest curves
give the peak numbers extrapolated to
the Northern (dotted lines) and Southern (dash-dotted) halves of the sky.
The remaining curves refer to the full sky.
Filled circles give the average numbers
of positive and negative peaks. The full line describes a power-law model with
$n = 1$ and  $Q_{%
{\rm rms-PS}} = 18.5$ $ \mu $K, corresponding to the fit
procedure S31 in Table 2}
\end{figure}
\begin{figure}
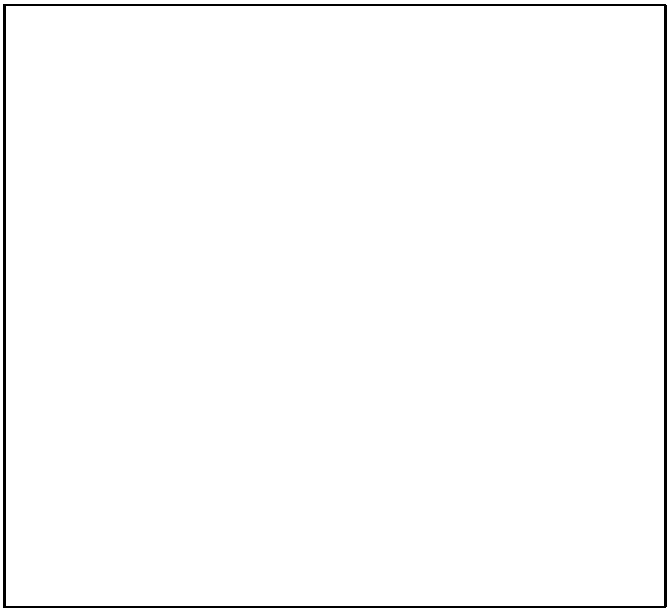

\picplace{8.0cm}
\caption[ ] {The noise harmonic coefficients
 $A_{{\rm N}\ell }^2$ (containing the form factors $W_\ell ^2$) and the beam
shape coefficients $G_\ell $ vs. $\ell.$
 Filled circles give $A_{{\rm N}\ell }^2$ as obtained from the noise map,
and the dotted line represents
the best fit curve corresponding to $\sigma _{\rm N}=0.936^{\circ }$
and $C_{\rm N}=5.74\times 10^{-5}$ mK$^2.$
Crosses give the experimental $G_\ell $, compared to Gaussian
approximations with
dispersion 2.9$^{\circ }$ (full line) and 3.1$^{\circ }$ (dash-dotted) }
\label{figure1}
\end{figure}

For the interpretation of the above data, we must consider that both cosmic
signal and noise contribute to the temperature field $T(\theta ,\phi )=\sum
a_{\ell m}Y_{\ell m}(\theta ,\phi )$. The expectation values of the numbers
of local maxima and minima are determined by the variances $a_\ell
^2=\sum_m\left\langle \mid a_{\ell m}\mid ^2\right\rangle $. For a Gaussian
field the full-sky $N_{{\rm peak}}$ is given by
\begin{eqnarray}
\label{Npeak} & &\left\langle N_{{\rm peak}}(\nu )\right\rangle  =
\sqrt{\frac 2\pi }
\theta^{*}{}^{-2}\left\{ \gamma ^{*2}\nu {\rm exp}\left( -\frac 12\nu ^2\right)
\right. \nonumber  \\
 & &+  \frac{\gamma ^{*}\left( 1-\gamma ^{*2}\right) ^{\frac 32}}{\sqrt{2\pi }}
{\rm %
exp}\left( -u_2^2\nu ^2\right) +\sqrt{\frac \pi 6}{\rm erfc}\left( \sqrt{%
\frac 32}u_1\nu \right) \nonumber \\
 & & -  \frac 12\int_\nu ^\infty \left[ \gamma ^{*2}(y^2-1){\rm %
exp}\left( -\frac 12y^2\right) {\rm erfc}\left( \gamma ^{*}u_2y\right) \right.
\nonumber \\
& &+  \left. \left. u_1%
{\rm exp}\left( -\frac 32u_1^2y^2\right) {\rm erfc}\left( \gamma
^{*}u_1u_2y\right) \right] {\rm d}y \right\} ,
\end{eqnarray}
where
\begin{equation}
\label{u1u2}u_1=\left( 3-2\gamma ^{*2}\right) ^{-\frac 12},\;\;\;u_2=\left[
2\left( 1-\gamma ^{*2}\right) \right] ^{-\frac 12},
\end{equation}
and the properties of the anisotropy field are summarized by the parameters $%
\theta ^{*}$ and $\gamma ^{*}$ (Bond \& Efstathiou 1987; Fabbri 1992) given
by
\begin{eqnarray}
\label{tetastar} \theta ^{*2} & = & \left[ 2\sum_\ell \ell \left( \ell
+1\right)
W_\ell ^2a_\ell ^2\right]  \nonumber \\
& \times & \left[  \sum_\ell \left( \ell -1\right) \ell \left(
\ell +1\right) \left( \ell +2\right) W_\ell ^2a_\ell ^2\right] ^{-1},
\end{eqnarray}
and
\begin{eqnarray}
\label{gammastar}\gamma ^{*2} & = & \left[ \sum_\ell \ell \left( \ell +1\right)
W_\ell ^2a_\ell ^2\right] ^2\left[ \sum_\ell W_\ell ^2a_\ell ^2\right]^{-1}
\nonumber \\
& \times & \left[ \sum_\ell \left( \ell -1\right) \ell \left( \ell +1\right)
\left( \ell +2\right) W_\ell ^2a_\ell ^2\right] ^{-1}.
\end{eqnarray}
Here $W_\ell $ are form factors taking into account beam shape and any
additional smearing effect.

Assuming that signal and noise are uncorrelated, their contributions to $%
a_\ell ^2$ add up in quadrature in the (A+B) maps, $a_\ell ^2=a_{{\rm S}\ell
}^2+a_{{\rm N}\ell }^2$, and we need to determine $a_{{\rm N}\ell }^2$ in an
independent way. This can be achieved by means of the (A$-$B) maps, which
can however be used to directly derive the coefficients $A_{{\rm N}\ell
}^2=W_\ell ^2a_{{\rm N}\ell }^2$ rather than the $a_{{\rm N}\ell }^2$.
Figure 2 reports the results of a harmonic best fit up to $\ell =30$
executed on the entire celestial sphere$.$ Increasing the number of
harmonics, best fits (as well as direct integration by means of $a_{\ell
m}=\int T(\theta ,\phi )Y_{\ell m}^{*}(\theta ,\phi ){\rm d}\Omega $) tend
to overrate large-$\ell $ amplitudes. We checked for this effect by
considering the peak statistics for noise maps.%
\begin{figure}
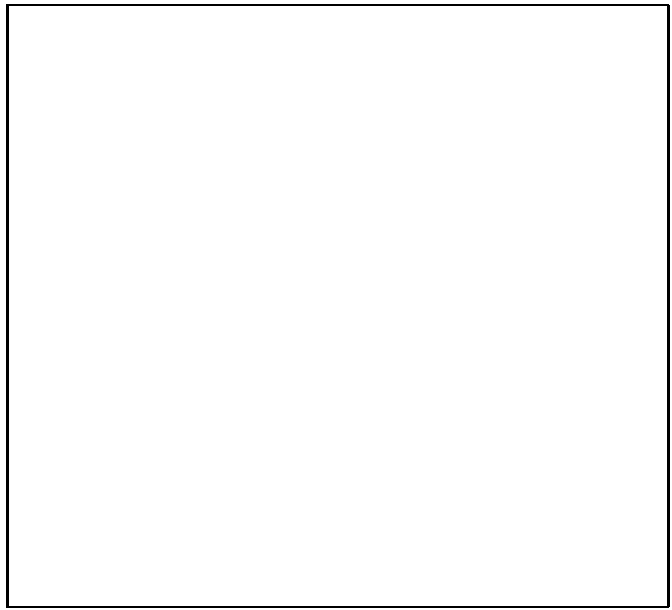

\picplace{8.0cm}
\caption[ ] {The peak number $ N_{\rm peak}$ vs. threshold in
COBE-DMR noise maps. Open
squares (triangles) denote positive (negative) peaks. The four lowest curves
give the peak numbers extrapolated to
the Northern (dotted lines) and Southern (dash-dotted) halves of the sky.
The remaining curves refer to the full sky. Filled circles give the
average statistics
of positive and negative peaks. The full line represents the best
fitting function}
\label{figure3}
\end{figure}
\begin{figure}
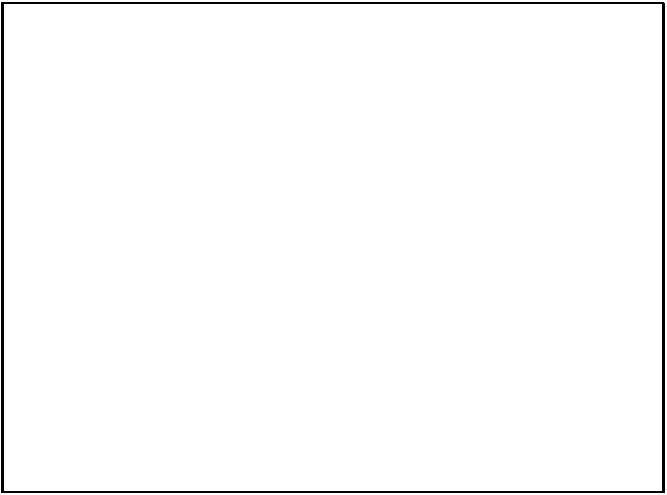

\picplace{6.5cm}
\caption[ ] {The acceptance regions in the $(n,Q_{{\rm rms-PS}})$ plane at 1-
and 2-sigma
confidence levels. Contours are computed with procedures S31 (full lines) and
 F31 (dotted). Both procedures assume
$\sigma_{\rm S} = 3.1^{\circ}$,
$\sigma_{\rm N} = 0.936^{\circ},$ and $C_{\rm N}
 =5.74 \cdot 10^{-5}  $ mK$^2$. The former uses the measured
$A_{\rm N\ell }^2$ up to $ \ell = 30$.}
\label{figure4}
\end{figure}
Figure 3 gives the numbers of maxima and minima detected in the full-sky
noise maps vs. threshold $\nu .$ If using Eq.s (\ref{Npeak}-\ref{gammastar})
we generate the peak statistics from a set of 50 harmonics with best-fitted
amplitudes, we get a low-threshold excess of about 30 peaks with respect to
the data in Fig. 2; the discrepancy increases with the number of harmonics.
Since accurate theoretical calculations of peak statistics require at least $%
\sim $50 harmonics, we tried to fit an analytic form to a more limited set
of $A_{{\rm N}\ell }^2$. A satisfactory choice is
\begin{equation}
\label{anoise}A_{{\rm N}\ell }^2=C_{{\rm N}}\left( \ell +\frac 12\right)
\exp \left[ -\left( \ell +\frac 12\right) ^2\sigma _{{\rm N}}^2\right] .
\end{equation}
However the large error bars on individual $A_{{\rm N}\ell }^2$ make a
2-parameter fit for the function (\ref{anoise}) not very useful. A more
convenient solution to this problem is to fit the parameter $\sigma _{{\rm N}%
}$ directly on the noise-generated peak distribution; using the average
distribution of maxima and minima (represented by the filled circles in
Figure 3) and 99 harmonics the result is $\sigma _{{\rm N}}=0.936^{\circ
}\pm 0.071^{\circ }.$ (We checked however that a set of $\sim 50\div 60$
harmonics would provide a sufficient accuracy.) The fit turns out to be even
``too'' good, providing $\chi _{\min }^2=1.3$ against 9 degrees of freedom:
This means that data points at different thresholds are not uncorrelated.
The value found for $\sigma _{{\rm N}}$ is quite independent of $C_{{\rm N}}$%
, since peak statistics does not depend on the overall amplitude of
anisotropies as shown by Eq.s (\ref{tetastar}, \ref{gammastar}).We then
fitted the amplitude parameter of Eq. (\ref{anoise}) on the reported set of $%
A_{{\rm N}\ell }^2$ with $\ell =2\div 30,$ getting $C_{{\rm N}}=(5.74\pm
0.66)\times 10^{-5}$ mK$^2.$ The distribution described bu the full line in
Fig. 3 was generated using the function (\ref{anoise}) in the entire range $%
\ell =2\div 99.$ We checked that the peak distribution of the (A-B) maps is
also recovered to a good accuracy building up the harmonic spectrum with the
``measured'' $A_{{\rm N}\ell }^2$ up to $\ell =30$ and the function (\ref
{anoise}) at higher $\ell .$

We tested power-law spectra against the peak statistics in Fig. 1 with the
following procedure. For each spectrum, labelled by the spectral index $n$
and the predicted quadrupole $Q_{{\rm rms-PS}}=a_{\rm S2}/\sqrt{4\pi },$ we
generated the theoretical expectation values $a_{{\rm S}\ell }^2$ and then
\begin{equation}
\label{totalAl}W_\ell ^2a_\ell ^2=G_\ell ^2\exp \left[ -\left( \ell +\frac
12\right) ^2\sigma _{{\rm S}}^2\right] a_{{\rm S}\ell }^2+A_{{\rm N}\ell }^2.
\end{equation}
Here $G_\ell $ denote the measured beam-shape coefficients of COBE-DMR.
These have reported by Wright et al. (1994) up to $\ell =50.$ A more
extensive set of $G_\ell $ up to $\ell =99$ has been provided to us by Kogut
(private communication). The exponential factor in Eq. (\ref{totalAl}) takes
into account the 2.9$^{\circ }$ smearing on the map as well as the
additional smearing due to orbital motion, so that our best estimate is $%
\sigma _{{\rm S}}=3.1^{\circ }$. Note that $\sigma _{{\rm S}}$ should not be
confused with the approximate Gaussian beamwidth which was used in many
computations, but not in the present work. Also, it is quite different from
the phenomenological parameter $\sigma _{{\rm N}}$ of Eq. (\ref{anoise}).
The peak distributions calculated from Eq.s (\ref{Npeak}-\ref{gammastar})
are tested against the averaged distribution of positive and negative peaks.
We should notice that such theoretical distribution are affected by several
sources of uncertainty. Cosmic variance (White et al. 1993) affects the
cosmic-structure contribution to the harmonic coefficients $a_{\ell m}$, and
a quite similar variance pertains to noise; in fact, these two effects are
described by identical equations assuming that both are Gaussian processes.
Performing a limited number of simulations for superpositions of Gaussian
signal and noise (with fixed $a_\ell ^2$, the {\em expectation} values of
harmonic strengths), we found that the probability distribution of $N_{{\rm %
peak}}$ at a given threshold is roughly consistent with a Poisson
distribution having a width $\left\langle N_{{\rm peak}}\right\rangle
^{\frac 12}.$ However, in our case the $A_{{\rm N}\ell }^2$ themselves are
not fixed. When we use Eq. (\ref{anoise}) their errors at different $\ell $
are correlated due to the uncertainties on $C_{{\rm N}}$ and $\sigma _{{\rm N%
}};$ it can be shown that the corresponding contribution to the uncertainty
on the predicted $N_{{\rm peak}}$ is given by
\begin{eqnarray}
\label{vars} & &{\rm var}(N_{{\rm peak}})  = \left( \sum_\ell A_{{\rm N}\ell
}^2%
\frac{\partial \left\langle N_{{\rm peak}}\right\rangle }{\partial A_{{\rm N}%
\ell }^2}\right) ^2\frac{{\rm var}(C_{{\rm N}})}{C_{{\rm N}}^2}
\nonumber \\
& & +4  \left[
\sum_\ell \left( \ell +\frac 12\right) ^2A_{{\rm N}\ell }^2\frac{\partial
\left\langle N_{{\rm peak}}\right\rangle }{\partial A_{{\rm N}\ell }^2}%
\right] ^2\sigma _{{\rm N}}^2{\rm var}(\sigma _{{\rm N}}).
\end{eqnarray}
When we use the measured $A_{{\rm N}\ell }^2,$ a more familiar equation $%
{\rm var}(N_{{\rm peak}})=\sum_\ell \left[ \left( \partial \left\langle N_{%
{\rm peak}}\right\rangle /\partial A_{{\rm N}\ell }^2\right) ^2{\rm var}(A_{%
{\rm N}\ell }^2)\right] $ applies. For each computed model we calculated the
corresponding $\chi ^2$ combining quadratically the above errors with with
the experimental error bars reported in Fig.1. This procedure allowed us to
avoid a more extensive use of simulations.

Figure 4 gives the allowed regions at 1 and 2 sigmas in the $(n,Q_{{\rm %
rms-PS}})$ plane. Full lines give the contours calculated for $\sigma _{{\rm %
S}}=3.1^{\circ },$ using the $A_{{\rm N}\ell }^2$ of Fig. 2 up to $\ell =30$%
, and the analytic form (\ref{anoise}) with the optimal values of $\sigma _{%
{\rm N}}$ and $C_{{\rm N}}$ at higher $\ell .$ (This procedure is referred
to as S31 in the Figure and in Table 2.) Contours calculated for $\sigma _{%
{\rm S}}=2.9^{\circ }$ (case S29, not reported in the Figure) would be
hardly distinguishable from them. Using the analytic form (\ref{anoise}) in
the entire range $\ell =1\div 99$ (case F31) the contours are slightly
displaced to smaller values of $Q_{{\rm rms-PS}}$ (by $\sim 2\ \mu $K for $%
n=1)$.
\begin{table}
  \caption[ ]{Coefficients of the $n-Q_{\rm rms-PS}$ regression defined by Eq.
(\ref{ab}) \label{regress}}
  \begin{flushleft}
  \begin{tabular}{lll}
  \hline \noalign {\smallskip}
     Procedure       &   $a$ ($\mu$K)          &  $b$ ($ \mu$K)   \\
  \noalign {\smallskip}  \hline  \noalign {\smallskip}
     S31$^\dagger$  &  $24.71\pm 0.65$       & $ -6.23\pm 0.45$ \\
     S29 $^\ddagger$    &  $24.54\pm 0.66$       & $ -6.16\pm 0.46$ \\
     F31$^{\star }$ &  $21.47\pm 0.60$       & $ -5.45\pm  0.42$ \\
  \noalign {\smallskip}  \hline
  \end{tabular}

 $^\dagger \sigma_{\rm N} =3.1^\circ$, using noise harmonic
 amplitudes from Fig. (2) for $\ell \leq 30$.

$^{\ddagger }\sigma_{\rm N} =2.9^\circ$

$^{\star }$Noise described by Eq. (\ref{anoise}) for any $\ell $
  \end {flushleft}
\end{table}

The parameters $n$ and $Q_{{\rm rms-PS}}$ are clearly anticorrelated, as
already found from analyses of harmonic amplitudes and the angular
correlation function (Seljak \& Bertschinger 1993; Torres et al. 1994; Smoot
et a. 1994). Minimizing $\chi ^2$ for fixed $n$ we identify a straight line
in the $(n,Q_{{\rm rms-PS}})$ plane,
\begin{equation}
\label{ab}Q_{{\rm rms-PS}}=a+bn,
\end{equation}
with the coefficient values listed in Table 2. From the above results,
taking into account differences arising from the S31 and F31 procedures, we
can conclude that $Q_{{\rm rms-PS}}=17\pm 3$ $\mu $K for $n=1$, and $Q_{{\rm %
rms-PS}}=14\pm 3$ $\mu $K for $n=1.5$. These numbers agree very well with
the most likely quadrupole $Q_{{\rm rms-PS}}$ derived from higher order
multipoles, but not with the quadrupole rms fluctuation of 6$\pm 3$ $\mu $K
directly fitted on two-year data; see Bennet et al. (1994) for a discussion
of this discrepancy, and Jing \& Fang (1994) for a possible explanation in
terms of an infrared cutoff in the spectrum.

\begin{acknowledgements}
We wish to thank A. Kogut for providing unpublished data on COBE-DMR
beam shape. This work is partially supported by Agenzia Spaziale
Italiana under Contract \# 94-RS-155, by the Italian Ministry for
University and Scientific and Technological Research (Progetti Nazionali e
di Rilevante Interesse per la Scienza), and by the European Union
under Contract \# CI1*-CT92-0013.
\end{acknowledgements}

\end{document}